\documentclass[12pt,preprint]{emulateapj}

\def\lsun{{\rm L_{\odot}}}

\def\msun{{\rm M_{\odot}}}

\def\Msun{\hbox{$\rm ~M_{\odot}$}}

\def\H0{{\rm ~km~s^{-1}~Mpc^{-1}}}

\def\msun{M_{\rm \odot}}

\def\go{
\mathrel{\raise.3ex\hbox{$>$}\mkern-14mu\lower0.6ex\hbox{$\sim$}}
}
\def\lo{
\mathrel{\raise.3ex\hbox{$<$}\mkern-14mu\lower0.6ex\hbox{$\sim$}}
}

\shorttitle{The SSS phase of RS Oph}
\shortauthors{J.P. Osborne et al.}

\begin{document}

\title{The Super-Soft X-ray phase of Nova RS Ophiuchi 2006}

\author{J.P. Osborne\altaffilmark{1}, K.L. Page\altaffilmark{1}, A.P. Beardmore\altaffilmark{1}, M.F. Bode\altaffilmark{2}, M.R. Goad\altaffilmark{1}, T.J. O'Brien\altaffilmark{3}, S. Starrfield\altaffilmark{4}, T. Rauch\altaffilmark{5}, J.-U. Ness\altaffilmark{6}, J. Krautter\altaffilmark{7}, G. Schwarz\altaffilmark{8}, D.N. Burrows\altaffilmark{9}, N. Gehrels\altaffilmark{10}, J.J. Drake\altaffilmark{11}, A. Evans\altaffilmark{12} and S.P.S. Eyres\altaffilmark{13}}

\altaffiltext{1}{Department of Physics and Astronomy, University of Leicester,
  Leicester, LE1 7RH, UK}
\altaffiltext{2}{Astrophysics Research Institute, Liverpool John Moores University, Birkenhead, CH41 1LD, UK}
\altaffiltext{3}{Jodrell Bank Observatory, School of Physics \& Astronomy, University of Manchester, Macclesfield, SK11 9DL, UK}
\altaffiltext{4}{School of Earth and Space Exploration, Arizona State University, P.O. Box 871404, Tempe, AZ 85287-1404, USA}
\altaffiltext{5}{Institute for Astronomy and Astrophysics, Kepler Center for Astro and Particle Physics, Eberhard Karls University, Sand 1, D-72076 T{\" u}bingen, Germany}
\altaffiltext{6}{XMM-Newton Science Operations Centre, ESAC, Apartado 78, 28691 Villanueva de la Ca{\~ n}ada, Madrid, Spain}
\altaffiltext{7}{Landessternwarte, K{\"o}nigstuhl, 69117 Heidelberg, Germany}
\altaffiltext{8}{American Astronomical Society, 2000 Florida Ave, NW, Suite 400, Washington, DC 20009-1231, USA}
\altaffiltext{9}{Department of Astronomy and Astrophysics, Pennsylvania State University, University Park, Pennsylvania, 16802, USA}
\altaffiltext{10}{NASA Goddard Space Flight Center, Greenbelt, Maryland 20771, USA}
\altaffiltext{11}{Smithsonian Astrophysical Observatory, 60 Garden St., MS 3, Cambridge, MA 02138, USA}
\altaffiltext{12}{Astrophysics Group, School of Physical and Geographical Sciences, Keele University, Staffordshire, ST5 5BG, UK}
\altaffiltext{13}{Jeremiah Horrocks Institute, University of Central Lancashire, Preston, PR1 2HE, UK}

\email{julo@star.le.ac.uk}

\begin{abstract}

  {\it Swift} X-ray observations of the $\sim$60~day super-soft phase
  of the recurrent nova RS Ophiuchi 2006 show the progress of nuclear
  burning on the white dwarf in exquisite detail.  First seen 26 days
  after the optical outburst, this phase started with extreme
  variability likely due to 
variable absorption, although intrinsic white dwarf variations are not 
excluded. About 32 days later, a steady decline in count-rate set in. 
NLTE model atmosphere spectral fits during the super-soft phase show that the 
effective temperature of the white dwarf increases from $\sim$65 eV to $\sim$90 eV during 
the extreme variability phase, falling slowly after about day 60 and more 
rapidly after day 80. The bolometric luminosity is seen to be approximately constant and close to 
Eddington from day 45 up to day 60, the subsequent decline possibly signalling 
the end of extensive nuclear burning.  Before the decline, a multiply-periodic,
  $\sim$35~s modulation of the soft X-rays was present and may be the
  signature of a nuclear fusion driven instability.  Our measurements
  are consistent with a white dwarf mass near the Chandrasekhar limit;
  combined with a deduced accumulation of mass transferred from its
  binary companion, this leads us to suggest RS Oph is a strong
  candidate for a future supernova explosion. The main uncertainty now
  is whether the WD is the CO type necessary for a SN Ia. This may
  be confirmed by detailed abundance analyses of
  spectroscopic data from the outbursts.
  \\

\end{abstract}

\keywords{novae --- binaries: symbiotic --- stars: oscillations --- X-rays: individual (RS Oph)}

\section{Introduction}

Novae result from the explosive thermonuclear fusion of hydrogen to
helium in the surface layers of a white dwarf (WD).  The hydrogen-rich
fuel for this process is provided by accretion from the outer layers
of a cooler binary companion (Starrfield 2008), in the case of RS Ophiuchi (RS~Oph),
a red giant (RG).  Most novae have a single historically recorded outburst
(the Classical Novae [CNe]), but a few, including RS Oph, are
recurrent on timescales of $\sim$8 to less than 80 years -- the so-called Recurrent Novae (RNe; see Schaefer 2010 for a review).  The observably
short recurrence intervals for RNe are thought to result
from a higher accretion rate and a higher WD mass than in CNe, both leading to a more rapid release of energy  in the ignition zone and a shorter time to runaway (Townsley 2008).

RS Oph shows recurrent nova outbursts at roughly 20 year
intervals. The latest outburst was detected on 2006 Feb 12.8 at a
magnitude of 4.5 (Narumi et al. 2006; see Hounsell et al. 2010 for a complete early light curve).  Multi-frequency observations
during the previous outburst in 1985 led to determinations of the
distance, d~=~1.6~$\pm$~0.3~kpc (Bode 1987), and the interstellar
column density,
N$_{\rm H,ISM}$~=~(2.4~$\pm$~0.6)~$\times$~10$^{21}$~cm$^{-2}$ (Hjellming
et al. 1986). X-ray observations were conducted by {\it EXOSAT} at six
epochs, from days 55 to 251 after the outburst, over the 0.04-2.0 keV
and 1.5-15 keV bands (Mason et al. 1987; O'Brien et al. 1992).  X-ray
emission during the first five epochs was consistent with shocks
propagating through the pre-existing wind of the RG companion
star. Modelling this process, O'Brien et al. (1992) derived an
outburst energy of 1.1~$\times$~10$^{43}$~erg and an ejected mass of
1.1~$\times$~10$^{-6}$~$\Msun$, but it proved difficult
to model the low and high energy X-ray evolution simultaneously and the authors
suggested that on-going nuclear fusion on the WD surface might be
responsible. As will be shown in this paper, our results confirm this view.

Novae have been predicted to undergo a super-soft source (SSS) phase
as the mass loss from the central source declines and the effective
photospheric surface shrinks at constant bolometric luminosity
(MacDonald et al. 1985).  This phase, which produces emission from the
surface of the WD with temperatures around 3~$\times$~10$^5$~K, has been
observed in several novae; see, e.g., Page et al. (2010); Drake et al. (2003); Ness et al. (2003);  Orio et
al. (2002).  Previously, the outburst of V1974 Cyg in 1992 had
the best temporal coverage, with 18 epochs of {\it ROSAT} observations
ranging from 63 to 653 days after outburst (Krautter et al. 1996,
Balman et al. 1998).  This outburst was the brightest SSS observed at the time (in terms of observed flux), but our {\it Swift} data show that RS~Oph peaked 2-3 times brighter
still.

X-ray data from the 2006 outburst of RS~Oph were obtained by {\it Swift}, the Rossi X-Ray Timing Explorer ({\it RXTE}), {\it XMM-Newton} and {\it Chandra} and have been extensively discussed in a range of papers. Bode et al. (2006) described the early hard emission, based on {\it Swift}-X-ray Telescope (XRT) observations (as well as a detection by the Burst Alert Telescope -- the hard X-ray instrument onboard {\it Swift}), confirming basic models from the 1985 outburst, while Sokoloski et al. (2006) presented the {\it RXTE} data for a similar early time interval. Ness et al. (2007, 2009) and Nelson et al. (2008) discuss the grating spectra obtained by {\it XMM-Newton} and {\it Chandra}, both before and during the SSS phase, while Drake et al. (2009) concentrated on the early (pre-SSS) {\it Chandra} high-energy transmission grating alone, finding the ejecta may contain super-solar abundances. 
Luna et al. (2009) find evidence for extended soft X-ray emission using {\it Chandra} CCD data.
Vaytet, O'Brien \& Bode (2007) present 1-dimensional hydrodynamical models of the shocks within the interacting winds of the RS~Oph system and compare them to the results of Bode et al. (2006). Their models reproduce the rise and subsequent deceleration of the shock velocities, but require a very high speed wind to achieve the high shock velocity observed. Three-dimensional modelling performed by Walder, Folini \& Shore (2008) leads them to conclude that the WD mass in the RS Oph system is increasing with time. Similar conclusions were reached by Orlando et al. (2009), who estimated a total ejecta mass of $10^{-6} M_\odot$ based on 3-D hydrodynamical modelling constrained by {\it Swift} and {\it Chandra} X-ray observations.

Besides the X-ray observations, data on RS~Oph were also collected in the radio (O'Brien et al. 2006; Kantharia et al. 2007; Eyres et al. 2009), IR (Monnier et al. 2006; Das, Banerjee \& Ashok 2006; Evans et al. 2007a,b; Chesneau et al. 2007; Lane et al. 2007; Banerjee, Das \& Ashok 2009; Rushton et al. 2010; Brandi et al. 2009) and optical bands (Hachisu et al. 2006; Worters et al. 2007; Bode et al. 2007; Munari et al. 2007; Brandi et al. 2009).

Hachisu, Kato \& Luna (2007) discussed the temporal evolution of the RS~Oph SSS
X-ray light curve, but did not include any detailed
consideration of the spectral characteristics or their evolution; nor
did they include the effects of interstellar and (changing)
circumstellar absorption on their conclusions. Here we perform
spectral fits to derive physical parameters such as effective
temperatures and luminosities, compare them to observational results
from other wavebands and also to theoretical expectations. In
addition, we explore temporal variability of the SSS on timescales
down to seconds and discuss the origin and implications of our
results. Finally, Hachisu et al. (2007) concluded that this system is a SN Ia
progenitor. Here we reconsider this important conclusion and suggest
additional analysis that is crucial to confirming or
rejecting this hypothesis.

\section{Swift XRT observations and data reduction}
\label{xrt}

The {\it Swift} satellite (Gehrels et al. 2004) is a rapid-response
observatory, designed for the study of Gamma-Ray Bursts (GRBs). It includes a
Burst Alert Telescope (BAT, 15--350 keV; Barthelmy et al. 2005), the XRT
(0.3--10 keV; Burrows et al. 2005), and a
UV-optical telescope (UVOT, 170-650 nm; Roming et al. 2005). {\it Swift}
observations started within 3.2 days of the outburst (defined as 2006
Feb 12.8) and were repeated in over 350 snap-shots between days 3 and
1565; they were scheduled according to the behaviour of the source,
although GRBs occasionally interrupted. 
The observations consisted of short
continuous snap-shots of 20--2600~s, and were made at multiples of the
{\it Swift} orbital period of 96 min. Observations were initially made at
2--8~days spacing up to 2006 March 13 (day 29). In response to an
observed dramatic count rate increase (Osborne et al. 2006a), the
observation frequency was increased to one per day until 2006 March 16 (day
32), and then to many a day (up to a maximum of 18) until the end of
April (day 77), after which it was reduced to around one observation
every three days. From 2006 May 24 (day 100), this was then further reduced to between
once a week and once a fortnight until the end of the observing window on 2006
October 22 (day 252). A further four observations were taken between days 372
and 392 (2007 February and March) when RS~Oph became observable again by {\it
Swift}, followed by $\sim$10~ks between days 592--596 (end of
September/beginning of October 2007), before the target again became
observationally constrained. $\sim$12.5~ks were collected between days 818-820
(May 2008), $\sim$13.3 ks between 2009 August and September and a final short observation was made in 2010 May.
The total exposure time of
the observations reported here is almost 450~ks. The complete {\it Swift} X-ray dataset to date is plotted in Fig.~\ref{fulllc}.

Before day 90 observations were made in Windowed Timing (WT) mode,
which provides one-dimensional spatial information and 1.8~ms time
resolution. This mode was selected because the alternative, Photon
Counting (PC) mode, with its longer integration time of 2.5~s, would
have resulted in unusably piled-up data -- pile-up occurs when more
than one photon provides charge to a pixel before readout. WT event
grades 0-2 were selected for analysis (Burrows et al. 2005), as this
selection gives good efficiency and discrimination against background,
and the event lists were cleaned in the standard manner using the {\it xrtpipeline}\footnote{Part of the XRT Data Analysis Software (XRTDAS) developed under the responsibility of the ASI Science Data Center (ASDC), Italy.} within {\it
  Swift} software version 3.4 (corresponding to HEASoft 6.7). The source data were
extracted using an annulus of inner and outer radii equal to five and 30
pixels (one XRT pixel = 2.36''). 
They were corrected for
pile-up by normalising for the excluded central pixels (which removed up to
75\% of the observed counts), and similarly for the presence of bad
CCD pixels. For the light-curve, this was done through the use of the {\it Swift} {\sc ftool} {\it xrtlccorr}, while the ancillary response files created using {\it xrtmkarf} performed the same function for the spectra.

By day 90 RS~Oph had faded to the point where PC mode
could be used; again we used a 30 pixel outer-radius source extraction
region. PC observations before day 100 were affected by pile-up and a
central exclusion region that varied from six down to zero pixels in
radius was applied to select only unaffected events.  PC event grades 0--12 were accumulated and corrected for the effects of excluded and bad
pixels, again using {\it xrtlccorr} and {\it xrtmkarf} where appropriate. The most up-to-date version of the calibration files (version 11 of the RMFs) was used.

\section{X-ray light-curve and spectrum}
\label{x}

After an initial modest increase in brightness in the second
observation (Figs~\ref{fulllc} and {\ref{lc-hr}), the X-ray count rate of RS Oph slowly fell to 6~count~s$^{-1}$ on day 26. During this initial interval, the spectrum was relatively hard, characterised by the emission expected from an optically thin plasma; these early observations are described by Bode et al. (2006). 
The following observation on day 29 showed RS~Oph at 15~count~s$^{-1}$, beginning a phase of dramatic increase in count rate and
large amplitude variations which peaked above 250~count~s$^{-1}$ (Osborne et
al. 2006b); Fig. 2 shows a change from 260 to 20~count~s$^{-1}$ in $\sim$12 hours during day 38. Most of the additional count rate resulted from a huge
increase at energies below 0.7~keV.  Around day 46 the rate
stabilised, peaking at more than 300~count~s$^{-1}$, then around day 58 a roughly linear decline began
(Osborne et al. 2006c). The X-ray emission (E~$<$~1~keV) seen in the
period from day~26 to day~58 was very much softer than the optically
thin spectrum, with kT $\sim$ a few keV, which had been seen earlier in the
outburst (Fig.~\ref{spec}).  The decline of the soft X-ray spectrum continued
to about day 90, when the SSS component was no longer evident (Osborne
et al. 2006d); the hardness of the spectrum, as measured by ratio of
the count rates (0.6-2.0 keV/0.3-0.6 keV), started to increase again
at this time, a trend which continued to day $\sim$120 (Fig.~\ref{lc-hr}), after which the hardness ratio stayed approximately constant.

\begin{figure}
\begin{center}
\includegraphics[clip,angle=-90,width=8cm]{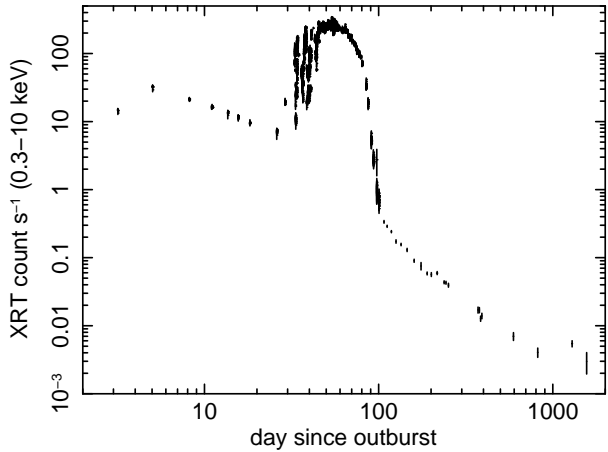}
\caption{The entire 0.3--10 keV {\it Swift}-X-ray light-curve of RS~Oph. The pile-up corrected count rate as a function of the number of days since outburst is plotted on log-log scales. The super-soft phase is prominent between days 29 and 100.}
\label{fulllc}
\end{center}
\end{figure}

\begin{figure*}
\begin{center}
\includegraphics[clip,angle=-90,width=12cm]{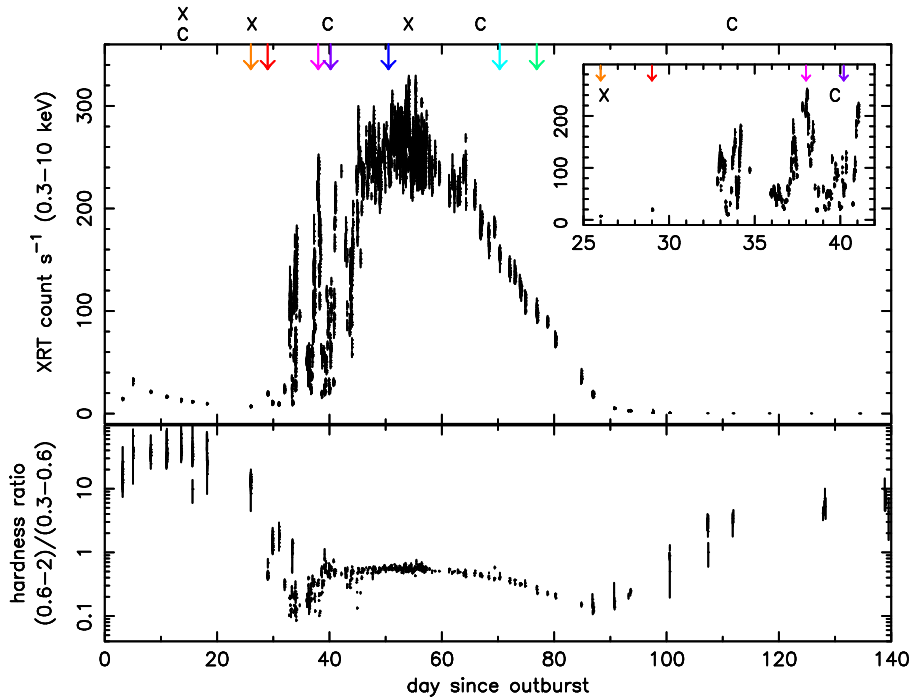}
\caption{The {\it Swift}-XRT light-curve (top) and the (0.6-2 keV)/(0.3-0.6 keV) count rate hardness ratio (bottom) for RS~Oph from 3 to 135 days after the outburst discovery.  The times of  high spectral resolution grating observations by {\it XMM-Newton} (X) and {\it Chandra} (C) are marked (e.g., Ness et al. 2007), as are the times of the XRT spectra (coloured arrows) shown in Fig.~\ref{spec}.
}
\label{lc-hr}
\end{center}
\end{figure*}

As in CNe, this X-ray behaviour is in stark contrast to that in the
optical, in which a relatively smooth decline in magnitude over
$\sim$140 days from the time of outburst was observed, albeit with a
break in slope around day 70 (as shown by the AAVSO dataset\footnote{http://www.aavso.org/}). While
the declining X-ray flux prior to day~26 is well described by the
evolution of shock systems established as the high-velocity ejecta
impacted the RG wind (Bode et al. 2006; Sokoloski et al. 2006),
the rapidly-appearing, bright and highly variable soft component
cannot be explained by this mechanism. The very high luminosity (see below) rules
out re-established accretion as the origin of this component.

The harder, shock-generated emission has a faster power-law decline after the SSS phase than before it. The maximum count rates due to this emission are below the pile-up limits for the modes used, so the light curve reflects the intrinsic behaviour of the source. Faster fading around the time of the SSS phase is suggested by the 1-D cooling model of Vaytet et al. (2007 \& 2010) which show that shock breakout of the red giant wind established since the previous nova explosion can occur then, and can reproduce the faster power-law decline. However, 3-D modelling is likely to be required for a complete understanding given the complex spatial structure of the observed remnant (Bode et al. 2007, Sokoloski, Rupen \& Mioduszewski 2008, Luna et al. 2009); current 3-D models (Walder, Folini \& Shore 2008, Orlando, Drake \& Laming 2009) unfortunately do not extend to this epoch.

High-resolution X-ray grating spectra from {\it Chandra} and {\it XMM-Newton}
(Ness et al. 2007, 2009) show that the soft component was dominated by
continuum emission over 0.3-0.8~keV on days 39, 53 and 66, with both
emission and absorption lines superimposed, indicating a hot WD
stellar atmosphere possibly with outflow. Emission features included lines attributable to the presence of H- and He-like N and H-like O.

\begin{figure}
\begin{center}
\includegraphics[clip,angle=-90,width=8cm]{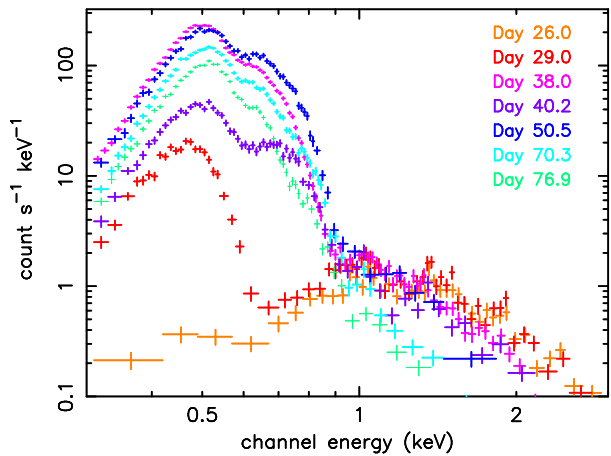}
\caption{The evolution of the {\it Swift} X-ray spectrum of RS~Oph through the outburst. Spectra from various spectral and intensity states are shown with the time after outburst. The spectrum on day~26 shows broad-band emission from the shocked gas and the first detection of a new low-energy component below 0.6~keV, while the day 29 spectrum clearly shows this super-soft source dominating the lower energies. The next two spectra come from the highly variable phase, and show differing flux fractions in the 0.6-1.0~keV band. By day~50, the flux from RS~Oph was more stable; the spectra from days 70.3 and 76.9 are taken from the decline phase. 
}
\label{spec}
\end{center}
\end{figure}

A sample of {\it Swift}-XRT spectra is shown in Fig.~\ref{spec} to demonstrate the large variation in spectral shape during the supersoft phase; the spectral evolution is discussed in more detail below. We note that the resolution of the XRT at 0.5 keV is only 63 eV (full width half maximum) and definitive detection of sharp spectral features requires the use of instruments with higher spectral resolution.  Nevertheless, the narrow emission and absorption features due to N and O seen in the grating data by Ness et al. (2007) are also discernible in the Swift-XRT spectra.

We characterised the SSS spectrum by minimum $\chi^2$ fits
to the observed 0.3-10~keV spectra using {\sc xspec} v12\footnote{http://heasarc.gsfc.nasa.gov/docs/xanadu/xspec/index.html}. Our model
consisted of a plane-parallel, non-local thermodynamic equilibrium atmosphere component (Rauch 2003; Rauch et al. 2010) and a two-temperature optically-thin solar abundance plasma to characterise the shocked wind, all absorbed by intervening gas (the
column density being the sum of the interstellar and the declining
absorption from the unshocked RG wind [Bode et al. 2006]). The model atmosphere grids were calculated for temperatures of (5.5 -- 10.5)~$\times$~10$^5$~K, in steps of 10$^4$~K. The declining column was constrained to follow a monotonic power-law decrease over time, of
the form N$_{\rm H,wind}$~=~7.5~$\times$~10$^{21}$~t$^{-0.5}$~cm$^{-2}$ (determined from Bode
et al. 2006, their fig.~4), where $t$ is the time since outburst in days. 
The two-temperature model for the optically thin
component is justified in each case by the large decrease in $\chi^2$
resulting from the inclusion of the second component (addition of a
third temperature component does not result in a significantly better
fit to any of the spectra); however it is not the  purpose of this paper to describe
the optically thin shock emission in detail (see Bode et al. 2008; Nelson et al. 2008; Ness et al. 2009). 

Analysis of the grating spectra mentioned above
(Ness et al. 2007) indicates that neutral oxygen is underabundant in
the circumstellar (RG wind) column: $\sim$30\% solar around day 40 and 0\%
by day 54. These values were therefore incorporated into the fits, switching
from 30\% O{\sc i} before day 54 to 0\% at later times as a simple approximation of the temporal change. 

Spectra were extracted for individual orbits of data, between days 26 and 100, to cover the vast majority of the SSS phase.
A script was written to fit the spectra
automatically within {\sc xspec}. 
A procedure was defined
such that the spectral fits were `shaken' out of local minima wherever
possible and the parameters further refined. Despite this, the fits
were often statistically poor, with $\chi^2_\nu$~$>$~2, because of the high statistical quality of the data and likely indicating the need for more sophistication in the physical model. 

Figure~\ref{sssfits} plots the results of the spectral fitting process. The
second panel shows how the temperature of the atmosphere model varies over time, starting from the point when the very soft emission was first detected. The third and fourth panels show the radius of the emitting region and the bolometric luminosity, respectively, both estimated from the model atmosphere. The uncertainties on the radius and luminosity are large, typically of order 10--20\% (90\% confidence) for the radius, and $\sim$40\% for the luminosity. The trend changes of the parameters are much larger than the errors on the individual values.

Although the underlying continuum was fairly well approximated by the Rauch atmosphere model, not all of the apparent emission features were accounted for. The Rauch models span abundances with respect to the Sun of carbon = 0.03--0.42 and nitrogen = 64--1.4, reflecting the expectation of CNO processing of accreted material. We used models 003, 004, 007 and 011\footnote{http://astro.uni-tuebingen.de/\raisebox{.2em}{\tiny
$\sim$}rauch/TMAF/\\flux\_HHeCNONeMgSiS\_gen.html.
In the framework of the Virtual Observatory ({\it VO}; http://www.ivoa.net),
these spectral energy distributions (SEDs, $\lambda - F_\lambda$)  are
available
in {\it VO} compliant form via the {\it VO} service  {\it TheoSSA}
(http://vo.ari.uni-heidelberg.de/ssatr-0.01/TrSpectra.jsp?)
provided by the {\it German Astrophysical Virtual Observatory}
({\it GAVO}; http://www.g-vo.org).}, but our fits did not indicate a preference between them. Figure~\ref{sssfits} shows the results from fits to model 003 which has
a C/N ratio of 4.8~$\times$~10$^{-4}$ (i.e., model B of Rauch et al. 2010). 
The main difference between the fits to the various abundance models was the temperature of the atmosphere after the rapid variability phase (model 011 gave a temperature $\sim$6\% lower than that from model 003, and a more gradual cooling after day 60); the best-fit bolometric luminosities were consistent with each other. 
Nelson et al. (2008) gave a preliminary temperature estimate of $\sim$8.2~$\times$~10$^5$~K ($\sim$71~eV) for the grating data obtained on day 54, compared to our value of $\sim$88~eV. Their value was based on a statistically poor eyeball fit, and this, together with differences in assumed abundances and intrinsic changes in RS~Oph, may account for the difference.

The 0.3--10~keV observed (unabsorbed) flux on day~50, around the time of the peak count-rate, was $\sim$5~$\times$~10$^{-9}$ (7~$\times$~10$^{-8}$) erg~cm$^{-2}$~s$^{-1}$, using the Rauch 003 model. This corresponds to a observed band-limited luminosity of $\sim$1.5~$\times$~10$^{36}$~erg~s$^{-1}$ (the majority of the total luminosity occurs in the unobservable band below 0.3 keV). The observed flux is completely dominated by the SSS emission: the optically thin components only provide approximately 1\% of the flux measured.

Following the interval of high-amplitude variability (see Section~\ref{rapid} for a discussion of these data),
the temperature of the SSS increases slightly (also shown by the
variation of the hardness ratio in Figure~\ref{lc-hr}). This continues
until around day~58--60 after outburst, at which point the temperature of the model atmosphere
starts to decrease slowly, following the count rate; the cooling becomes more rapid after $\sim$day~80. The increase in
temperature during the early supersoft phase is predicted theoretically (Starrfield et al. 1991; Balman et
al. 1998) and should continue until nuclear burning ceases.

Starrfield et al. (1991) gives the radius of a 1.4~M$\msun$ WD as 1.9~$\times$~10$^8$~cm, which is approximately the minimum effective radius for the SSS measured here, indicating the material has probably all fallen back to an undisturbed WD radius by day 100.

\begin{figure}
\begin{center}
\includegraphics[clip, width=8cm]{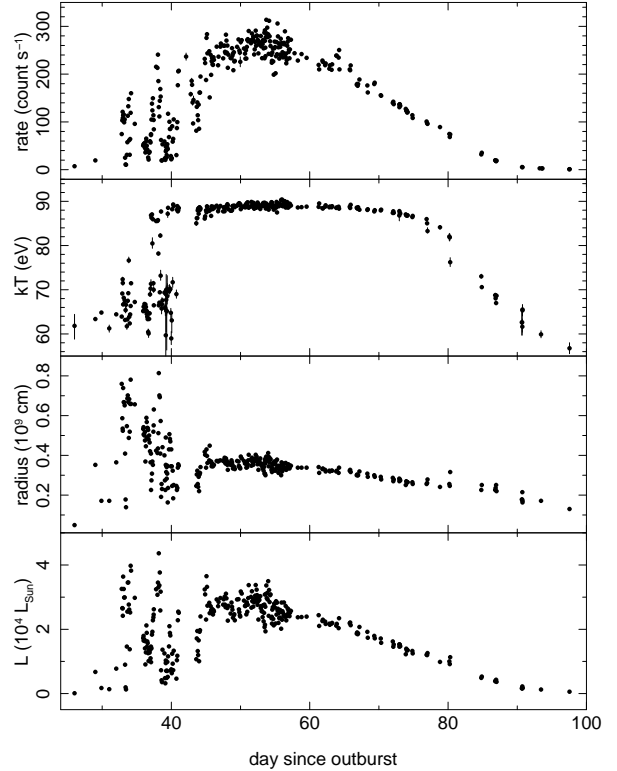}
\caption{Chronological sequence of {\sc xspec} fits of the Rauch atmosphere model grid 003 to the {\it Swift}-XRT SSS spectra of RS~Oph:  temperature (second panel), effective radius (third panel) and bolometric luminosity (fourth panel). 30\% oxygen was used until the start of day 54, 0\%
  afterwards. The light-curve (top panel) has been binned to have a single point per snapshot. No error bars have been plotted for the radius or luminosity; see text for discussion.}
\label{sssfits}
\end{center}
\end{figure}

In the past, blackbodies have frequently been used as an approximation to the emission spectrum when fitting supersoft sources. However, such fits can lead to an underestimate of the temperature and an overestimate of the luminosity (e.g., Krautter et al. 1996). Fitting our RS~Oph  data with blackbodies gave peak temperatures during the `plateau' (between days $\sim$40 and 60) of about 65~eV, with a corresponding effective radius of $\sim$7~$\times$~10$^8$~cm and a bolometric luminosity of around 3~$\times$~10$^4$~$\lsun$, compared to $\sim$90~eV, 4~$\times$~10$^8$~cm and 3~$\times$~10$^4$~$\lsun$ derived from the model atmosphere fits. Surprisingly, the plateau blackbody and model atmosphere luminosities are the same; however during the high-amplitude variability phase the blackbody luminosity peaked at 20--30~$\times$~10$^4$~$\lsun$, much higher than the model atmosphere peak. The model atmosphere trends of an initial slight increase in temperature from around day 44 to 55, then a cooling after day 60 or so, together with a short approximately constant luminosity  phase followed by fading, were also seen from the blackbody results. Preliminary blackbody fits to the RS~Oph data were discussed by Page et al. (2008).

We can compare the luminosities we derive during the SSS phase with
other results. Snijders (1987) derived an observed initial luminosity
peak for the 1985 outburst about four times the Eddington luminosity (L$_{\rm Edd}$,
the critical luminosity above which radiation pressure exceeds local
gravity; this is 4.8~$\times$~10$^4$~$\lsun$ for a 1.4~$\msun$ WD). There was then a `plateau' at
$\sim$L$_{\rm Edd}$ for days 8--35 and it took $\sim$57 days for the bolometric
luminosity to drop by around a factor of two. Iben and Tutukov (1996) give
the theoretical relationship of the bolometric luminosity to WD mass
during the constant bolometric luminosity (plateau) phase as L$_{\rm P}$~$\sim$~4.6~$\times$~10$^4$(M$_{\rm WD}$$-$0.26)$\lsun$. For M$_{\rm WD}$~=~1.4$\msun$, this gives L$_{\rm P}$~$\sim$~5.2~$\times$~10$^4$~$\lsun$. The luminosities we find from our spectral fits ($\go$~2--3~$\times$~10$^4$~$\lsun$ during the almost-constant count-rate phase, ie $\sim$L$_{\rm Edd}$) are in
gratifying agreement, and are consistent with a high WD mass.

The appearance of the SSS around one month after outburst is predicted
by assuming both a constant bolometric luminosity and that the peak
of the emission evolves to shorter wavelengths as the effective
photospheric radius decreases (caused by the decreasing mass-loss rate
from the WD; MacDonald, Fujimoto and Truran, 1985). Bath and Harkness
(1989) give an expression relating the change in $V$ magnitude from
peak, $\Delta V$, to T$_{\rm eff}$ as
\begin{equation}
T_{eff} = T_0 \times 10^{\Delta V / 2.5} K
\end{equation}
where T$_0$~=~8000~K is the more recent accepted value (Evans et al. 2005,
and references therein). By t~=~30~days, $\Delta V$ ~$\approx$~4.6~mag\footnote{http://www.aavso.org/} and thus T$_{\rm eff}$~$\approx$~5.5$\times$~10$^5$~K ($\sim$45--50~eV) would be expected at this time. 
Although this is an
approximation (it does not take into account the contribution of
nebular emission in the $V$ band, and it assumes that the photosphere emits as a blackbody), the derived value of T$_{\rm eff}$ is in line
with our observations, helping to confirm our association of the onset
of the SSS phase in RS~Oph with the appearance of the nuclear burning
source in the XRT band.

\subsection{High-amplitude flux variability phase}
\label{rapid}

From shortly after the initial detection of the soft component until about 46~days after the outburst, the X-ray count rate varied dramatically, covering two orders of magnitude. Previous CNe have also shown variability during the super-soft phase.
For example, Drake et al. (2003) observed a
large, short-lived `flare' during the SSS phase of V1494~Aql. Similarly, Orio et al. (2002) observed a sudden transient flux
decrease in V382~Vel with a slight softening of the spectrum, and Ness
et al. (2003) found rapid variations and a `puzzling episodic
turn-off' accompanied by spectral softening in V4743~Sgr. 
This
behaviour has not yet been explained in the literature. The oscillations
cannot readily be ascribed to changes in absorption in neutral
intervening gas because of their spectral signature; neither is this behaviour
reproduced by existing nova simulations, which do not show variations
on these timescales (Starrfield 2008). 

While it remains unclear if the variability seen in RS~Oph has the
same origin as the other cases cited here, it seems likely that our
dense observations of RS Oph have discovered a new phenomenon. Since this initial discovery (Osborne et al. 2006b), V458 Vul (Ness et al. 2009), Nova LMC 2009a (Bode et al. 2009; Bode et al. in prep), V2491~Cyg (Ness et al. 2008; Ness et al. 2010) and KT Eri (Beardmore et al. 2010) have all revealed large amplitude variability in their soft X-ray light-curves, remarkably similar to the RS Oph results. One
possibility is that clumpiness in the expanding ejected WD envelope
could cause variable, and possibly ionized, absorption of the soft
X-ray flux between the early total X-ray extinction and the late
optically thin phases, broadly consistent with what we have seen in RS
Oph. This is investigated in more detail in this section.

Figure~\ref{brightsoft} shows a short section of the {\it Swift}-XRT light-curve and hardness ratio. During this time, the emission is typically softer when the source is brighter (see top plot). However, the lower panel highlights snapshots during which the emission hardens as the count rate increases. 

\begin{figure}
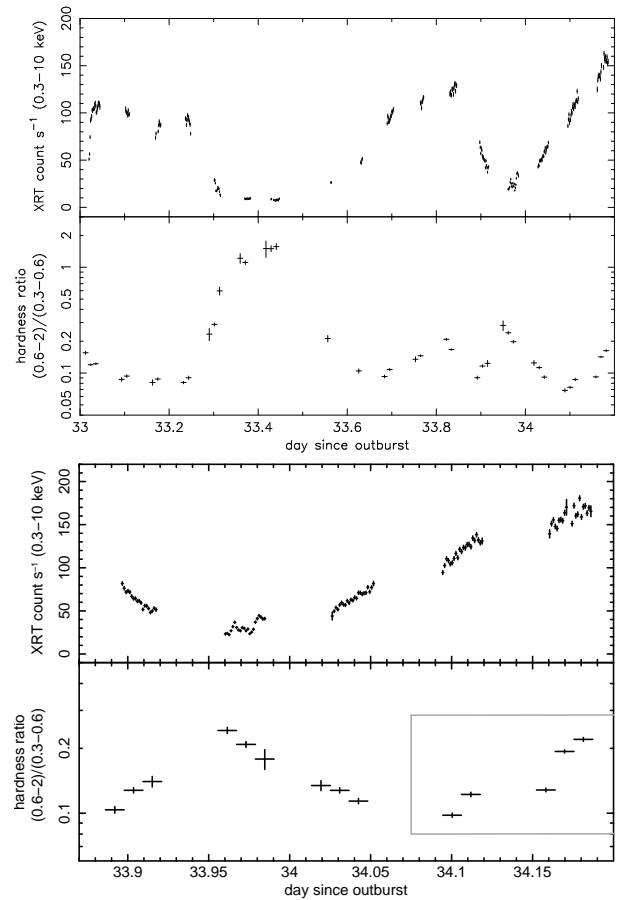

\begin{center}
\includegraphics[clip, angle=-90, width=8cm]{brightsoftnew.ps}
\includegraphics[clip, angle=-90, width=8cm]{brighthardzoom.ps}
\caption{The top plot shows the X-ray light-curve and hardness ratio between days 33 and 34.2 after outburst. It can be seen that the general trend is for the X-rays to be softer when the emission is brighter. However, the lower plot zooms in on days 33.9--34.2, demonstrating that some of the snapshots (in the grey box) become harder as the count rate increases.}
\label{brightsoft}
\end{center}
\end{figure}

\begin{figure}
\begin{center}
\includegraphics[clip, width=9cm]{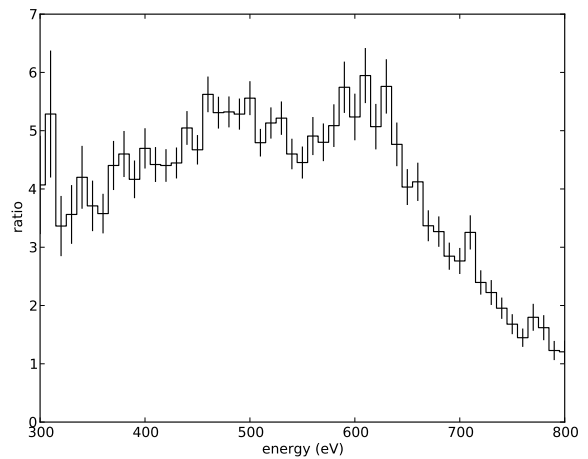}
\caption{The ratio of observed counts per channel spectra from a `peak' and a `trough' of the high amplitude variability phase. The spectal shape of this ratio rules out simple variations in cold absorption or photospheric radius as the cause of the observed flux variations. The spectral resolution is 63 eV FWHM at 500~eV.}
\label{ratio}
\end{center}
\end{figure}

The most frequently observed trend of being harder at times of low flux is consistent with variable visibility of the hot white dwarf, at low flux the relatively constant hotter optically thin emission contributes a larger fraction of the observed counts. 
Photospheric radius variation or variable neutral absorption can in principle account for this behaviour. However a varying photospheric temperature may also be implicated: at low temperatures there is little soft X-ray flux in the XRT band, while the soft count rate increases sharply with higher temperatures so lowering the hardness; at the highest temperatures photospheric flux can appear in the harder spectral band causing the hardness to rise. In addition to these possibilities, variable ionization of the absorbing medium may play a role, allowing softer flux to escape when it is more highly ionized.
 
Figure~\ref{ratio} shows the ratio of a `bright' and a `faint' spectrum during this interval of variability and demonstrates that we are probably seeing a combination of effects. It appears to rule out both a variable neutral absorption and a radius change as neither would show a declining ratio to lower energies. Variations in the ionization state of realistic ionized wind absorption also do not show a declining spectral ratio to lower energies. An increased photospheric temperature when brighter does produce a spectral ratio qualitatively similar to that of Fig.~\ref{ratio}, especially in combination with decreased radius, but we have been unable to find a good quantatative match to this ratio using our fitted model atmospheres and shock models. 

If the absorption is changing in some complex fashion during this phase of the evolution, then constraining the column to decay as N$_{\rm H,wind}$~=~7.5~$\times$~10$^{21}$~t$^{-0.5}$~cm$^{-2}$ (Section~\ref{x}) may be an inappropriate approximation. Repeating the fits with a varying N$_{\rm H,wind}$ presented the same overall trends, however. 

High resolution X-ray grating spectra can be valuable in diagnosing the large flux modulation (e.g. see sect. 3.5 of Ness et al. 20007), although a persuasive physical explanation for this remains to be found.

\section{Rapid quasi-periodic variability}

A strong short-period modulation of the soft X-ray flux was seen
during much of the SSS phase (Osborne et al. 2006b). This modulation
was first detected on day~32.9 and continued until last detected on
day 58.8, coincident with the onset of the flux decline (see
Fig.~\ref{35s}). Soft X-ray light curves in the band 0.3-1~keV were created
with 1~s binning, with a duration of 1024 seconds and at least 75\%
exposure for all WT mode observations (i.e. up to day 90) in order to
characterise this modulation via Fourier transforms. We verified by simulations that the oscillations
in the X-ray flux were not strictly periodic, ruling out rigid rotation of
the WD as the origin. Although the excess Fourier power in the entire
dataset was centred on 35.3~s, detected periods formed a distribution with FWHM points at 34.5 and
38.5~s before day~49, but only between the adjacent frequency bins
corresponding to periods of 34.7 and 35.9~s after that date (see top panel of Fig.~\ref{period}). Periods and coherence were measured by fits to the power spectra;
we found P~=~35.82~$\pm$~0.06~s for all the data before day~49 and P~=~34.88~$\pm$~0.02~s after; the 1$\sigma$ ranges quoted here reflect the
distribution of power rather than measurement error; the coherence of
the signal is a factor of three poorer in the earlier data. The modulation
amplitude varied up to $\sim$10\%; neither the modulation amplitude nor the period were correlated with intensity (Beardmore et al. 2008). The lower panel of Fig.~\ref{period} illustrates the modulation during $\sim$830~s on day 37.2, when it was particularly strong.
The modulation was primarily sinusoidal, occasionally showing
a first harmonic; there is no detectable spectral variation.
Comparison of
spectral fits using the 003 model atmosphere to spectra accumulated over 0.2 cycle phase intervals at the phases of
maximum and minimum flux did not show temperature changes above
2\%. More details are provided by Beardmore et al (2008), who also
illustrate the 35~s flux variation in the light-curve and the
simultaneous presence of multiple periods\footnote{The current paper updates the statements regarding spectral variation in this reference.}. The 35~s modulation was
also seen in contemporaneous {\it XMM-Newton} data (Ness et al. 2007).

\begin{figure}
\begin{center}
\includegraphics[clip,width=9.5cm]{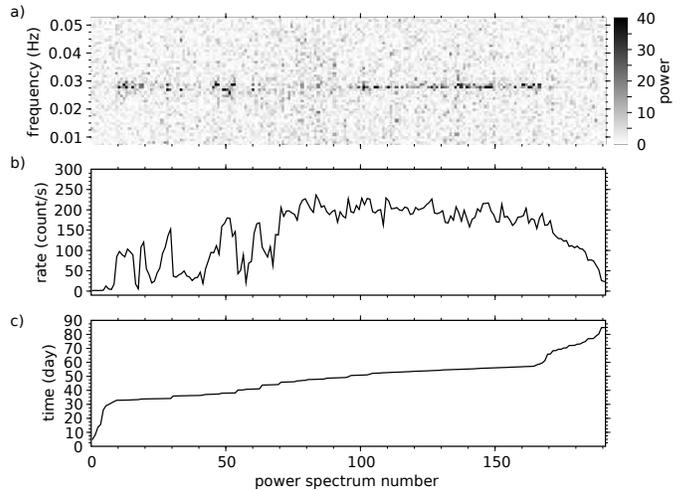}
\caption{Oscillatory modulation of the soft X-ray flux of RS Oph. a) Fourier power spectra for successive observations; each power spectrum is constructed from a 1024 s light curve of 1 second binning in the 0.3-1.0~keV band and the unbinned Fourier power is shown on a linear grey scale. b) The source intensity in each power spectrum. c) The relation between time and the power spectrum index number (used as the x-axis of each plot); time does not increase uniformly as we have suppressed times when there were no observations. Sporadic power can be seen between frequencies of 0.026 and 0.029~Hz, corresponding to periods of 34.5 to 38.5~s. While the modulation is generally present when RS~Oph is bright, counter-examples can be seen around bins 33 (day 36.0) and 45 (day 37.2), and the modulation is no longer present after bin 167 (day 58.8).
}
\label{35s}
\end{center}
\end{figure}

\begin{figure}
\begin{center}
\includegraphics[clip,width=9.2cm]{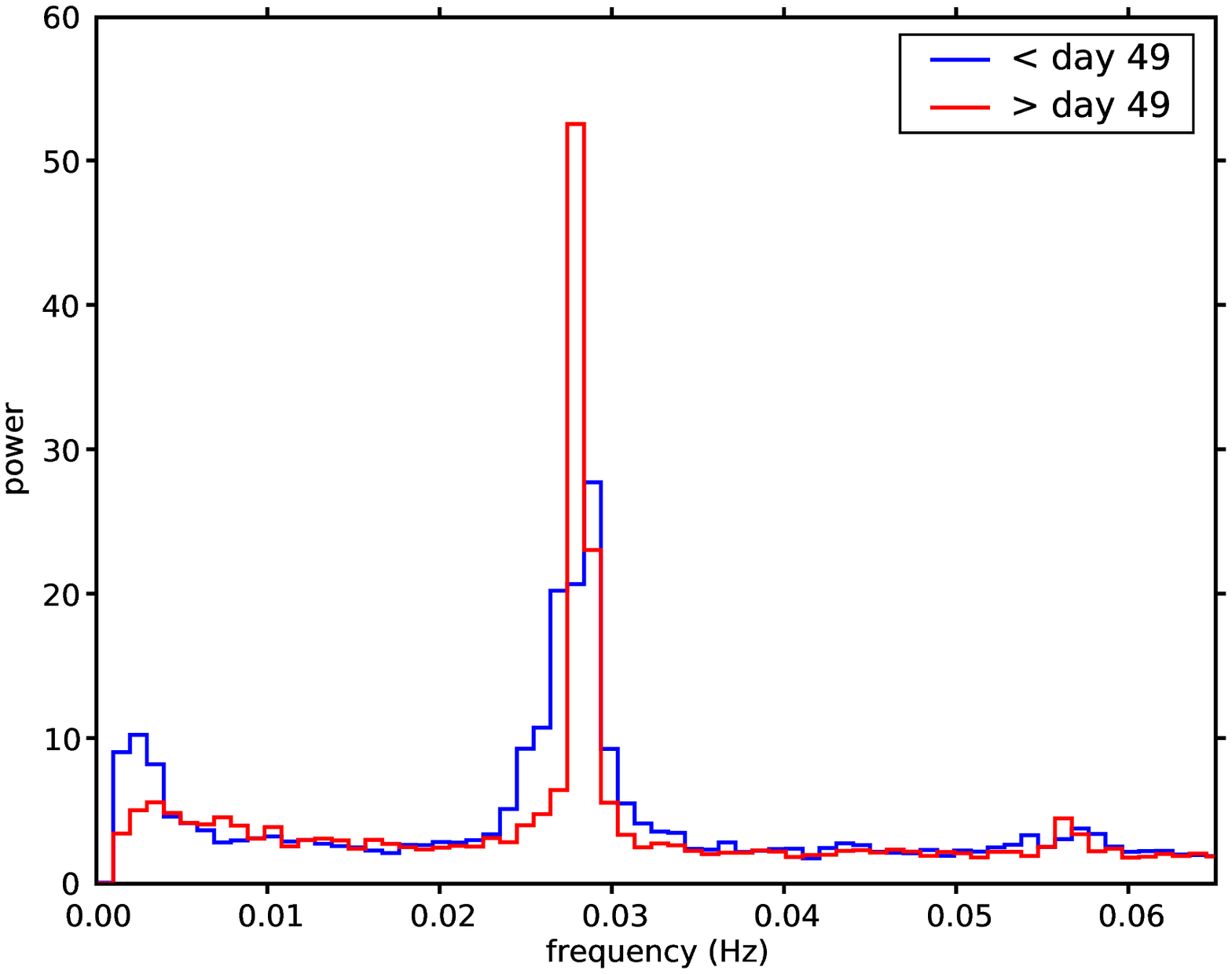}
\includegraphics[clip,width=6cm,angle=-90]{018f_0.3-10_34.0475_fld.ps}
\caption{The $\sim$35 s quasi-period of RS Oph. Top plot: Fourier power spectra of the aggregated light curves used in Fig.~\ref{35s}, showing the increased coherence of the 35 s modulation after day 49. A weak harmonic can be seen at ~0.056 Hz; aperiodic red noise is visible at the lowest frequencies. Bottom plot: Normalised soft X-ray light curve from day~37.2, when the modulation was particularly strong, folded at the period then present. 
}
\label{period}
\end{center}
\end{figure}

This $\sim$35~s periodicity is by far the shortest period seen in the SSS
phase of a nova to-date, although periodic oscillations have been
observed in other novae during this phase (P~=~22 min in V4743~Sgr, Ness
et al. 2003; P~=~42 min amongst others in V1494~Aql, Drake et
al. 2003). A period of 38~min has also been seen in the X-ray flux of
the persistent super-soft source CAL~83 (Schmidtke \& Cowley
2006). These longer period oscillations have been ascribed to g+-mode
(buoyancy) pulsations driven by an ionisation-opacity instability such
as occurs in pulsating planetary nebulae nuclei (PNN) or GW~Vir stars
(Drake et al. 2003). It has however been noted (Drake et al. 2003) that
the so-called $\epsilon$-mechanism could be responsible
and this should produce much shorter periods (these are oscillations in which increased nuclear energy generation is followed by expansion leading to a reduced generation rate which is followed by contraction which leads to an increased rate); such oscillations could be manifest at the SSS photosphere as cyclic changes in radius and or temperature.
Kawaler (1988) finds
periods of 70--200~s from this mechanism for a 0.62~$\msun$ PNN and one
expects the period, P, to decrease with the mass of the WD
(e.g. if P~$\propto$~$\rho_m^{-0.5}$, where $\rho_m$ is the mean density of the WD from
Starrfield et al. [1991], then M$_{\rm WD}$~$\sim$~0.9--1.25$\msun$). Such oscillations might
therefore be the source of the modulation we see in RS~Oph in the SSS
phase: they would only last as long as the nuclear burning source was
still active on the WD surface, as is observed. This reinforces the
proposition that nuclear burning is the origin of the SSS emission
prior to the decline, at which point the oscillations in RS Oph cease,
if the decline is due to the cessation of widespread nuclear fusion. Since the
models of Kawaler relate to L~$<$~1600$\lsun$ and have long instability
growth times, further work is required to establish whether the
$\epsilon$-mechanism is indeed responsible for the short period modulation we
have seen. If this turns out to be the case, then we can expect strong
constraints on the mass and structure of the WD.

We consider that Dwarf Nova Oscillations (DNO), which typically have
periods $\sim$30 s, are unlikely to be related to the modulations we have
detected. Ascribed to the accretion disk-WD boundary layer in dwarf
novae and some nova-like binaries, models of DNO depend on a form of
magnetically controlled accretion loosely tied to the body of the WD
(Warner 2004). The high luminosity of RS~Oph during the SSS phase
rules out an accretion-related origin of the 35-s period. Also ruled
out is an intermediate polar mechanism, since the same luminosity
argument applies, and the period changes seen are not consistent with
realistic torques on the WD.

{\it Swift} observations of KT~Eri (Nova Eri 2009) also revealed a $\sim$35~s modulation (Beardmore et al. 2010), with fractional amplitudes of up to 7\%; as in RS~Oph the periodicity was not continuously present. The existence of a very similar modulation in two different novae adds weight to the tentative conclusion that this period does not have a rotation-based origin, and the surprising similarity of the (very short) periods might suggest that both objects contain WDs near the Chandrasekhar limit.

\section{Mass and luminosity in the RS Oph system}

In this section, we obtain
estimates of the fundamental parameters for RS Oph. The timing of the
start of the final decline (i.e., day~58) is an important
parameter. The turn-off time for nuclear burning, t$_{\rm rem}$, is a steep
function of WD mass; MacDonald (1996) finds, for example, t$_{\rm rem}$~$\propto$~M$_{\rm WD}^{-6.3}$. The fact that the deduced nuclear burning timescale in RS~Oph is
much shorter than that observed in other novae (e.g. V1974~Cyg, where
t$_{\rm rem}$~$\gtrsim$~511~days and M$_{\rm WD}$~$\sim$~1.2$\msun$ [Krautter et al. 1996; Balman et
al. 1998]) is thus consistent with the existence of a much higher mass
WD in RS~Oph. Furthermore, from Starrfield et al. (1991), the observed
turn-off timescale in RS~Oph implies M$_{\rm WD}$~$\sim$~1.35$\msun$ (see also Hachisu
et al. 2007).

The critical mass of the accreted envelope required to trigger a thermonuclear runaway is given by

\begin{equation}
M_{acc} \approx \frac{4\pi R_{WD}^4}{G M_{WD}} P_{crit}	
\end{equation}
(Truran \& Livio 1986), where  P$_{\rm crit}$~$\sim$~10$^{20}$~dyne~cm$^{-2}$ for M$_{\rm WD}$~=~1.4$\msun$. Thus with R$_{\rm WD}$~=~1.9~$\times$~10$^8$~cm (Starrfield et al. 1991), M$_{\rm acc}$~$\approx$~4.4~$\times$~10$^{-6}$~$\msun$. This estimate of the accreted mass necessary for outburst is in line with the results of numerical simulations of the recurrent nova outburst using solar abundance material (Starrfield et al. 1988) and the estimate of Hachisu et al. (2007). We may compare this to the mass burnt, M$_b$, from

\begin{equation}
M_b = \frac{E_{SE} + E_p + E_{LO} + \Delta E_{KE} - E_{CE}}{X \epsilon_H}	
\end{equation}
where E$_{\rm SE}$ is the luminous energy in the initial
super-Eddington phase with luminosity L$_{\rm SE}$ lasting t$_{\rm
  SE}$ days; E$_{\rm p}$~=~L$_{\rm p}$(t$_{\rm rem}$$-$t$_{\rm SE}$) is
the luminous energy in the constant luminosity `plateau' phase;
E$_{\rm LO}$~=~(GM$_{\rm WD}$$\Delta$M)/R$_{\rm WD}$ is the energy
required to lift off the ejected mass, $\Delta$M; $\Delta$E$_{\rm KE}$
is any additional kinetic energy given to the ejecta; E$_{\rm CE}$ is
the energy imparted to the expanding envelope by the secondary star in
the so-called common envelope stage where the ejecta envelop the
secondary; $X$ is the mass fraction of hydrogen and
$\epsilon_H$~=~6.4~$\times$~10$^{18}$~erg~g$^{-1}$.

From the observational and theoretical considerations of the nuclear
burning phase given above, we take L$_{\rm p}$~=~2.5~$\times$~10$^4$~$\lsun$ and t$_{\rm rem}$~=~58~days. With
L$_{\rm SE}$~$\sim$~2~$\times$~10$^5$~$\lsun$ and t$_{\rm SE}$~$\sim$7~days, derived from the 1985 outburst, E$_{\rm SE}$~=~4.8~$\times$~10$^{44}$~erg and E$_{\rm
  p}$~=~2.5~$\times$~10$^{44}$~erg. Using the WD parameters above,
E$_{\rm LO}$~=~(2--20)~$\times$~10$^{44}$~erg for
$\Delta$M~=~10$^{-7}$--10$^{-6}$$\msun$ (Hjellming et al. 1986;
O'Brien et al. 1992; Sokoloski et al. 2006). From the results of the
shocked wind model fits to the 1985 outburst data (O'Brien et
al. 1992), $\Delta$E$_{\rm KE}$ appears negligible in
comparison (although more recent modelling by Vaytet et al (2010) yields energies more comparable to those from the plateau, see also Orlando et al. 2009).
E$_{\rm CE}$ is insignificant for this wide binary. We therefore
conclude that M$_b$~=~(1--3)~$\times$~10$^{-7}$$\msun$ (for $X$~=~0.7, since the matter from the RG will be relatively H-rich). Thus the mass burnt is a few percent of the mass
of the accreted envelope.

We now turn to the monotonic decline following the plateau
phase. Krautter et al. (1996) suggested that the decline they saw in
V1974~Cyg was consistent with energy radiated from gravitational
contraction of the extended WD atmosphere, comprising material not
ejected at the time of outburst, once the nuclear burning source was
extinguished. A gradual softening of the source was also noted, as
seen in RS Oph at this stage (see Fig.~\ref{lc-hr}), consistent with cooling of
the emitting region. From consideration of the Kelvin-Helmholtz
timescale, $\tau_{\rm K}$, associated with the contraction of this envelope, mass
M$_{\rm A}$, it can be shown that 

\begin{equation}
M_A = \left(\frac{4R_{WD}}{3GM_{WD}}\right) L_p \tau_K  
\end{equation}

Thus with $\tau_{\rm K}$~$\sim$~1~month, and L$_{\rm p}$, M$_{\rm WD}$
  and R$_{\rm WD}$ as above, M$_{\rm
    A}$~$\sim$~1~$\times$~10$^{-7}$$\msun$. Both M$_{\rm b}$ and
  M$_{\rm A}$ are less than M$_{\rm acc}$ as expected, and our value
  of M$_{\rm A}$ is much less than $\sim$10$^{-5}$$\msun$ found for
  the classical nova V1974~Cyg (Krautter et al. 1996) -- again
  implying that the WD in RS~Oph is much more massive than the
  $\sim$1.2$\msun$ WD in V1974~Cyg.

With the accreted mass between outbursts M$_{\rm acc}$~$\approx$~4.4~$\times$~10$^{-6}$$\msun$ and the
ejected mass at outburst $\Delta$M~=~10$^{-7}$--10$^{-6}$$\msun$, we confirm the
conclusion of Hachisu et al. (2007) that matter is gradually accumulating on
the extremely high mass WD in RS~Oph. Thus, it seems that the WD is
growing in mass to the Chandrasekhar limit, leading to its ultimate
demise as a (possible) Type Ia supernova (as also predicted for U Sco by Starrfield et al. 1988). 

Marietta et al. (2000) have
shown that entrainment of secondary star hydrogen during the supernova
explosion may reveal itself as low velocity lines becoming visible in
the late stages of an explosion which, however, have never been seen
in a Type Ia explosion. Hachisu et al. (2007) also conclude that
RS~Oph will explode in this way, based on the same {\it Swift}
observations as presented here. Against this, the recent work of Hayden et al. (2010) shows that red giants are disfavoured as the companions of SN Ia progenitors.  Another major uncertainty is the nature of the
WD; if the WD is originally of the ONe type rather than CO, and matter
is added to take it beyond the Chandrasekhar limit, an implosion will
occur rather than a supernova explosion. Thus the WD type is crucial
to determining the ultimate fate of the RS~Oph system. The key here
may come from abundance analyses of spectroscopy taken both before and
during the outburst. This is discussed in more detail in terms of the
results of high resolution X-ray spectroscopy from {\it Chandra} and {\it XMM-Newton} by
Ness et al. (2009).

Finally, from M$_{\rm acc}$ and the inter-outburst (1985 to 2006)
period, we derive a mean accretion rate $\dot {M}_{\rm
  acc}$~=~2~$\times$~10$^{-7}$~$\msun$~yr$^{-1}$. We note that this is
very similar to the net mass-loss rate in the red giant wind escaping
the system derived in Hjellming et al. (1986: $\dot {M}_{\rm
    RG}$~=~8~$\times$~10$^{-8}$--2.4~$\times$~10$^{-7}$~$\msun$~yr$^{-1}$
  for v$_{\rm wind}$~=~10--20~km~s$^{-1}$), but much lower than the
  1.8~$\times$~10$^{-6}$~$\msun$~yr$^{-1}$ derived by Bohigas et
  al. (1989). Using

\begin{equation}
L_{acc} = (1-\alpha) \frac{G M_{WD} \dot {M}_{\rm acc}}{2R_{WD}}	
\end{equation}
where $\alpha$~=~0.5 (Starrfield et al. 1988; see also Popham \& Narayan 1985), we find a mean observable inter-outburst accretion luminosity, L$_{\rm acc}$~=~3.1~$\times$~10$^{36}$~erg~s$^{-1}$. Spectral fits to our {\it Swift} data at the very latest epochs give L$_{\rm X}$~$\sim$~few~$\times$~10$^{31}$~erg~s$^{-1}$, suggesting that accretion is not yet fully re-established. This {\it Swift} measurement is of the same order as the X-ray luminosity of (3--20)~$\times$~10$^{31}$~erg~s$^{-1}$ derived from {\it ROSAT} observations some years after the 1985 outburst (Orio 1993), with the flux from our late-time spectral fit being of the same order of magnitude as the upper bound of the {\it ROSAT} estimate. Such a low value has yet to be explained in our understanding of the evolution of RS~Oph, this subject is examined by Mukai (2008) who speculates that the mass-accumulating white dwarf would be highly spun up, thus suppressing the boundary layer X-ray luminosity. In any case, to determine L$_{\rm acc}$ accurately, observations at lower energies (UV and optical) are also needed. Indeed, we note that Snijders (1987) found a much higher accretion luminosity, around that required, from analysis of IUE observations during quiescence of RS~Oph before the 1985 outburst.

\section{Summary}
\label{conc}

Our {\it Swift} X-ray observations of the 2006 outburst of the recurrent
nova RS~Oph have shown it to be the brightest SSS yet observed, and
have followed the progress of the nuclear burning source on the
surface of the white dwarf in unprecedented detail. First seen on day
26, the super-soft phase started with a phase of extreme variability;
the flux stabilised around day~46 at around $L_{Edd}$ and began a decline around day~58. This high-amplitude variability may be a product of both varying temperature and `blobby', possibly ionized, absorption. During days~33--59, a $\sim$35~s modulation was seen, which became increasingly
coherent through this interval. 

From these observations we have
confirmed the basic model of the outburst and, more importantly,
concluded that this may be a progenitor system for a future supernova
event, the main uncertainty being the nature of the white dwarf. However, our discoveries pose further observational and
theoretical challenges. These include understanding the nature of the
short period oscillations and the early extreme variability uncovered
by our observations, as well as confirming observationally the mass
accretion rate we predict to be required to give rise to the next
explosion.

\section{Acknowledgments}

We thank the {\it Swift} Science Team and the {\it Swift} Mission Operations Team for their excellent support of this observing campaign. JO, KP, MG, AB, MFB acknowledge the support of STFC. SS acknowledges partial support to ASU from NSF and NASA. TR is supported by the German Aerospace Center (DLR) under grant
05\,OR\,0806. The authors are grateful to John Nousek for his helpful comments on an initial draft of this paper and to Simon Vaughan for providing the TCL `shakefit' procedure.

\end{document}